\documentstyle[12pt]{article}
\newcommand\be{\begin{equation}}
\newcommand\ee{\end{equation}}
\newcommand\bea{\begin{eqnarray}}
\newcommand\eea{\end{eqnarray}}

\newcommand{\fatalpha}{{\bf \alpha \kern -0.44em \alpha}}
\newcommand{\fatsigma}{{\bf \sigma \kern -0.54em \sigma}}
\newcommand{\tpchi}{{\bf \chi \kern -0.35em \chi}}
\newcommand{\llambda}{{\bf \lambda \kern -0.45em \lambda}}

\bibliography{plain}
\title{\bf Entanglement entropy and Schmidt numbers in quantum networks of coupled quantum oscillator}\vspace{20mm}
\author{ M. A. Jafarizadeh$^{a}$
\thanks{E-mail:jafarizadeh@tabrizu.ac.ir},
 S. Nami$^{a}$
 \thanks{E-mail:S.Nami@tabrizu.ac.ir}
 F. Eghbalifam$^{a}$
 \thanks{E-mail:F.Egbali@tabrizu.ac.ir},
\\ $^a${\small Department of Theoretical Physics and Astrophysics,
University of Tabriz, Tabriz 51664, Iran.}} \pagebreak

\vspace{20mm}
\begin{document}
\maketitle \vspace{15mm}

\begin{abstract}
We investigate the entanglement of
the ground state in the quantum networks that their nodes are
considered as quantum harmonic oscillators. To this aim, the Schmidt numbers and
entanglement entropy between two arbitrary partitions of a
network, are calculated.

In partitioning an arbitrary graph into two parts there are some nodes in each parts which are not connected to the nodes of the other part. So these nodes of each part, can be in distinct subsets. Therefore the graph separates into four subsets. The nodes of the first and last subsets are those which are not connected to the nodes of other part.
In theorem I, by using generalized Schur complement method in these four subsets, we prove that all graphs which their connections between all two alternative subsets are complete, have the same entropy. A large number of graphs satisfy this theorem. Then the entanglement entropy in the limit of large coupling and large size of system, is
investigated in these graphs.

One of important quantities about partitioning, is conductance of graph. The conductance of graph is considered in some various graphs. In these graphs we compare the conductance of graph and the entanglement entropy.

\end{abstract}

\newpage
\section{Introduction}
Entanglement is a quantum mechanical phenomenon in which particles that are arbitrary
distances apart can influence each other instantly.
The unique relationship between entangled particles makes them very useful
for information processing. with the advent of quantum
information theory, entanglement was recognized as a resource, enabling tasks like quantum cryptography [1], quantum
teleportation [2] or measurement based quantum computation [3].
Since entanglement has become regarded as such an important resource, there is a
need for a means of quantifying it.

For the case of bipartite entanglement, a recent exhaustive review was
written by the Horodecki family [4] and entanglement measures have been reviewed in detail by Virmani and Plenio
[5]. One of the operational entanglement criteria is the Schmidt
decomposition [6-8]. The Schmidt decomposition is a very good
tool to study entanglement of bipartite pure states. The Schmidt
number provides an important variable to classify entanglement.
The entanglement of a partly entangled pure state can be
naturally parametrized by its entropy of entanglement, defined as
the von Neumann entropy, or
equivalently as the Shannon entropy of the squares of the Schmidt
coefficients [6,8]. L. Amico et al [9], reviewed the properties of the entanglement in many-body systems.
An important class of works analyzing the entanglement in many body system considered a bipartition of
the system dividing it in two distinct regions $A$ and $B$. If the total system is in a pure state
then a measure of the entanglement between $A$ and $B$ is
given by the von Neumann entropy $S$ associated to the
reduced density matrix of one of the two blocks $\rho_A (\rho_B)$.
Dealing with higher dimensional space of the local degrees of freedom generally involves complications which
are not tamable within the current knowledge about entanglement. The Peres-Horodecki criterion, is not sufficient already for two three-level systems. The situation simplifies
if only so called \textit{Gaussian states} of the harmonic oscillator modes are considered [10-15]. With gaussian states, it has been possible to implement certain quantum tasks as quantum teleportation, quantum cryptography and quantum computation with fantastic experimental success. The importance of gaussian states is two-fold; firstly, its structural mathematical description makes them much more amenable than any other continuous variable system (Continuous variable systems are those described by canonical conjugated coordinates $x$  and $p$ endowed with infinite dimensional Hilbert spaces). Secondly, its production, manipulation and detection with current optical technology can be done with a very high degree of accuracy and control.
In [16] the authors quantified the amount of information that a
single element of a quantum network shares with the rest of the
system. They considered a network of quantum harmonic oscillators
and analyzed its ground state to compute the entropy of
entanglement that vacuum fluctuations creates between single
nodes and the rest of the network by using the Von Neumann
entropy.

In this work we quantify the entanglement entropy
between two parts of the network. To this aim, we compute the vacuum state of bosonic modes
harmonically coupled through the specific adjacency matrix
of a given network.
In the section II, first we describe the Hamiltonian of our
model in subsection 2.1. Also we demonstrate the Schmidt
decomposition and entanglement entropy in 2.2.

In section III, The entanglement entropy and the Schmidt numbers of the ground state wave
function between two parts of the network, are calculated. It is performed by using some
local transformations in three stage.

In the section IV, the generalized Schur complement method is used for calculating the Schmidt numbers and entanglement entropy. In partitioning a graph into two parts, an arbitrary graph with every kind of cut, can be shown as a graph with four subsets. This is performed by separating the nodes of each part into two subsets. In each part, the nodes which are connected to the nodes of the other part, constitute a subset and the nodes which are not connected to the nodes of other part, constitute the other subset. So the graph is composed from four subsets that the nodes of first and last subsets are those which don't have any connection with the nodes of other part (Fig II). We use the generalized Schur complement method in the potential matrices of graphs to eliminate the connection matrices of first and last subsets. Then one can calculate the entanglement entropy between two parts of graph.
We give a theorem about entanglement entropy in spacial graphs. In this theorem we prove that all graphs which their connections between all two alternative subsets are complete, have the same entropy. Then two corollaries are resulted from Theorem I.
In example I, the entanglement entropy between two parts of Complete graph is calculated analytically by using corollary I.
In example II, the entanglement entropy between two equal parts of path graph is calculated analytically by applying generalized Schur complement method.
In example III, the entanglement entropy of the Barbell graph is considered. It is an important example which satisfies the Theorem I.
In example IV, we calculate the entanglement entropy of lollipop graph which satisfies the corollary II.

In section V, the conductance of graph, is defined. Then the entanglement entropy and conductance of graph are compared in some kinds of graphs. Also from the amount of entanglement entropies in the examples of this section, we found that the entanglement entropy has direction relation with the number of connections between two parts. When two kinds of partitions have the same number of connections between two parts, the entanglement entropy has direction relation with the number of Schmidt number for those partitions.

In section VI, we investigate the entanglement entropy in the limit of large coupling and large size of system.

Finally in section VI, we compare the result of our method with previous work in the case that the entanglement entropy between one node and the rest of the system, is considered. We give the criticism for the formula of previous work [9].

\section{preliminaries}
\subsection{The model and Hamiltonian}
We consider nodes as identical quantum oscillators, interacting
as dictated by the network topology encoded in the Laplacian $L$.
The Laplacian of a network is defined from the Adjacency matrix
as $L_{ij} = k_i\delta_{ij}- A_{ij}$ , where $k_i =\sum_j A_{ij}$
is the connectivity of node $i$, i.e., the number of nodes
connected to $i$. The Hamiltonian of the quantum network thus
reads:
\begin{equation}
H=\frac{1}{2}(P^T P+ X^T(I+2gL)X)
\end{equation}
here $I$ is the $N \times N$ identity matrix, $g$ is the coupling
strength between connected oscillators while $p^T=(p_1,p_2,...,
p_N)$ and $x^T=(x_1,x_2, ..., x_N)$ are the operators
corresponding to the momenta and positions of nodes respectively,
satisfying the usual commutation relations: $[x, p^T] = i\hbar I$
(we set $\hbar = 1$ in the following) and the matrix $V=I+2gL$ is
the potential matrix. Then the ground state of this Hamiltonian
is:
\begin{equation}
\psi(X)=\frac{(det(I+2gL))^{1/4}}{\pi^{N/4}}exp(-\frac{1}{2}(X^T(I+2gL)X))
\end{equation}
Where the $A_g=\frac{(det(I+2gL))^{1/4}}{\pi^{N/4}}$ is the
normalization factor for wave function. The elements of the
potential matrix in terms of entries of adjacency matrix is
$$V_{ij}=(1+2g\kappa_i)\delta_{ij}-2gA_{ij}$$

\subsection{Schmidt decomposition and entanglement entropy}

Any bipartite pure state $|\psi\rangle_{AB} \in
\textsl{H}=\textsl{H}_A \otimes\textsl{H}_B$ can be decomposed, by
choosing an appropriate basis, as
\begin{equation}
|\psi\rangle_{AB}=\sum_{i=1}^m \alpha_i|a_i\rangle\otimes|b_i\rangle
\end{equation}
where $1 \leq m \leq min\{dim(\textsl{H}_A); dim(\textsl{H}_B)\}$,
and $\alpha_i
> 0$ with $\sum_{i=1}^m \alpha_i^2 = 1$. Here $|a_i\rangle$ ($|b_i\rangle$) form a part of an
orthonormal basis in $\textsl{H}_A$ ($\textsl{H}_B$). The positive
numbers $\alpha_i$ are called the Schmidt coefficients of
$|\psi\rangle_{AB}$ and the number $m$ is called the Schmidt rank
of $|\psi\rangle_{AB}$.

Entropy of entanglement is defined as
the von Neumann entropy of either $\rho_A$ or $\rho_B$:
\begin{equation}
E=-Tr\rho_A log_2\rho_A= Tr\rho_B log_2\rho_B=-\sum_i\alpha_i^2
log_2 \alpha_i^2
\end{equation}

\section{Entanglement entropy between two parts of a network}
we want to introduce a method to quantify the entanglement entropy between two arbitrary parts of a network.
first we divide the potential matrix of the system into two parts, So
the potential matrix ($I+2gL$) can be written in the form
\begin{equation}
\left(\begin{array}{cc}
          A& B\\
            B^T& C\\
          \end{array}\right)
\end{equation}
where the size of block $A$ is $m\times m$, $C$ is
$(N-m)\times(N-m)$ and $B$ is $m\times(N-m)$. We assume that the
vector $X$ be decomposed of two sets $X,Y$ (i.e.,
$X=(X|Y)=(x_1,x_2, ..., x_m,y_1,y_2,...,y_{N-m})$).

We know that any local operation dosen't change the entanglement
between the nodes, so we apply some of these operations to
calculate the entanglement entropy for different lattices easily.
First we want to diagonalize the blocks $A$ and $C$, to this aim
we apply the local rotations $O_A$ and $O_C$ to the matrix of
(3-5), resulting as:
\begin{equation}
\psi(\widehat{x},\widehat{y})=\frac{(det(I+2gL))^{1/4}}{\pi^{N/4}}exp(-\frac{1}{2}(\widehat{x}\quad
\quad \widehat{y})\left(\begin{array}{cc}
          D_A& O^{\dagger}_A B O_C\\
            O_C^{\dagger}B^T O_A& D_C\\
          \end{array}\right)\left(\begin{array}{c}
          \widehat{x}\\
            \widehat{y}\\
          \end{array}\right))
\end{equation}
where $$\widehat{x}=O^{\dagger}_A x$$
$$\widehat{y}=O_C^{\dagger} y$$ We can define
$$\hat{B}=O^{\dagger}_A B O_C$$

then the wave function becomes
\begin{equation}
\psi(\widehat{x},\widehat{y})=A_g
exp(-\frac{1}{2}(\widehat{x}\quad \quad
\widehat{y})\left(\begin{array}{cc}
          D_A& \hat{B}\\
            \hat{B}^T & D_C\\
          \end{array}\right)\left(\begin{array}{c}
          \widehat{x}\\
            \widehat{y}\\
          \end{array}\right))
\end{equation}
In the next stage, the blocks $D_A$ and $D_C$ can be transformed
to Identity matrices by rescaling the variables $\widehat{x}$ and
$\widehat{y}$:
$$\widetilde{x}=D_x^{1/2}\widehat{x}$$
$$\widetilde{y}=D_y^{1/2}\widehat{y}$$
so the ground state wave function is transformed to
\begin{equation}
\psi(\widetilde{x},\widetilde{y})=A_gexp(-\frac{1}{2}(\widetilde{x}\quad
\quad \widetilde{y})\left(\begin{array}{cc}
          I& \widetilde{B}\\
            \widetilde{B}^T& I\\
          \end{array}\right)\left(\begin{array}{c}
          \widetilde{x}\\
            \widetilde{y}\\
          \end{array}\right))
\end{equation}
where $\widetilde{B}=D_x^{-1/2}\hat{B}D_y^{-1/2}$

In the third stage, we should calculate the singular value
decomposition of matrix $\widetilde{B}$:
$$U\widetilde{B}V^{\dagger}=D_B$$
$$U\widetilde{x}=x'$$
$$V\widetilde{y}=y'$$
The ground state wave function after this local operation is:
\begin{equation}
\psi(x',y')=A_gexp(-\frac{1}{2}(x'\quad \quad y')
\left(\begin{array}{cc}
          U& \textbf{0}\\
            \textbf{0}& V\\
          \end{array}\right)
\left(\begin{array}{cc}
          I& \widetilde{B}\\
            \widetilde{B}^T& I\\
          \end{array}\right)\left(\begin{array}{cc}
          U^{\dagger}& \textbf{0}\\
            \textbf{0}& V^{\dagger}\\
          \end{array}\right)
          \left(\begin{array}{c}
          x'\\
            y'\\
          \end{array}\right))
\end{equation}

\begin{equation}
\psi(x',y')=A_gexp(-\frac{1}{2}(x'\quad \quad
y')\left(\begin{array}{cc}
          I& U\widetilde{B}V^{\dagger}\\
            V\widetilde{B}^TU^{\dagger}& I\\
          \end{array}\right)\left(\begin{array}{c}
          x'\\
            y'\\
          \end{array}\right))
\end{equation}
The singular value decomposition(SVD), transforms matrix
$\widetilde{B}$ to the diagonal matrix $D_B$, so diagonal blocks
of the matrix in equation (3-5) will be diagonal, then the final
form of wave function is:
$$\psi(q^{x}_1,q^{x}_2,...,q_m^{x},q_1^y,q_2^y,...,q_{N-m}^y)=$$
\begin{equation}
A_gexp(-\frac{1}{2}(q_1^x,q_2^x,...,q_m^x,q_1^y,q_2^y,...,q_{N-m}^y)\left(\begin{array}{cccccccc}
          1& 0&\ldots & 0& d_1& 0&\ldots & 0\\
          0& 1 & \ldots &0&0&d_2&\ldots &0\\
          \vdots &\vdots &\ddots &\vdots &\vdots &\vdots &\ddots &\vdots\\
          0&0&\ldots & 1&0&0&\ldots&d_m\\
          d_1& 0&\ldots & 0&1& 0&\ldots & 0\\
          0&d_2&\ldots &0&0& 1 & \ldots &0\\
          \vdots &\vdots &\ddots &\vdots &\vdots &\vdots &\ddots &\vdots\\
          0&0&\ldots&d_{N-m}&0&0&\ldots & 1\\
          \end{array}\right)\left(\begin{array}{c}
          q_1^x\\
            q_2^x\\
            \vdots \\
           q_m^x\\
           q_1^y \\
           q_2^y \\
           \vdots \\
           q_{N-m}^y \\
          \end{array}\right))
\end{equation}
The above equation can be written in the form
$$\psi(q^{x}_1,q^{x}_2,...,q_m^{x},q_1^y,q_2^y,...,q_{N-m}^y)=$$

$$A_ge^{-\frac{(q_1^x)^2}{2}-\frac{(q_1^y)^2}{2}-d_1q_1^xq_1^y}\times
e^{-\frac{(q_2^x)^2}{2}-\frac{(q_2^y)^2}{2}-d_2q_2^xq_2^y}\times
\ldots \times
e^{-\frac{(q_m^x)^2}{2}-\frac{(q_m^y)^2}{2}-d_mq_m^xq_m^y}$$
\begin{equation}
\times e^{-\frac{(q_{m+1}^y)^2}{2}}\times
e^{-\frac{(q_{m+2}^y)^2}{2}}\times \ldots \times
e^{-\frac{(q_{N-m}^y)^2}{2}}
\end{equation}
From above equation, it's clear that the node $q_i^x$ is just
entangled with $q_i^y$, so we can use following identity to
calculate the schmidt number of this wave function,
\begin{equation}
\frac{1}{\pi^{1/2}}exp(-\frac{1+t^2}{2(1-t^2)}((q_i^x)^2+(q_i^y)^2))+\frac{2t}{1-t^2}q_i^x
q_i^y)=(1-t^2)^{1/2}\sum_n t^n \psi_n(q_i^x)\psi_n(q_i^y)
\end{equation}
In order to calculating the entropy, we apply a change of
variable as
$$1-t^2=\frac{2}{\nu+1}$$
$$t^2=\frac{\nu-1}{\nu+1}$$
So the above identity becomes
\begin{equation}
\frac{1}{\pi^{1/2}}exp(-\frac{\nu}{2}((q_i^x)^2+(q_i^y)^2))+(\nu^2-1)^{1/2}q_i^x
q_i^y)=(\frac{2}{\nu+1})^{1/2}\sum_n (\frac{\nu-1}{\nu+1})^{n/2}
\psi_n(q_i^x)\psi_n(q_i^y)
\end{equation}
and the reduced density matrix is
\begin{equation}
\rho=\frac{2}{\nu+1}\sum_{n}(\frac{\nu-1}{\nu+1})^n |n\rangle
\langle n|
\end{equation}
the entropy is
\begin{equation}
S(\rho)=-\sum_n p_n log(p_n)
\end{equation}
where $p_n=\frac{2}{\nu+1}(\frac{\nu-1}{\nu+1})^n$
\begin{equation}
\sum_n p_n log(p_n)=log(\frac{2}{\nu+1})+\langle n \rangle
log(\frac{\nu-1}{\nu+1})
\end{equation}
and $\langle n \rangle = \frac{\nu -1}{2}$

\begin{equation}
S(\rho)=\frac{\nu +1}{2} log(\frac{\nu +1}{2})-\frac{\nu
-1}{2}log(\frac{\nu -1}{2})
\end{equation}
By comparing the wave function (3-12) and the identity (3-14) and
define the scale $\mu^2$, we conclude that
$$\nu_i=1 \times \mu^2$$
$$(\nu_i^2-1)^{1/2}=-d_i \times \mu^2$$
After some straightforward calculation we obtain
\begin{equation}
\nu_i=(\frac{1}{1-d_i^2})^{1/2}
\end{equation}
By above discussion we conclude that
$$e^{-\frac{(q_1^x)^2}{2}-\frac{(q_1^y)^2}{2}-d_1q_1^xq_1^y}=\sum_n \lambda_{1,n}\psi_n(q^x_1)\psi_n(q^y_1)$$
$$e^{-\frac{(q_2^x)^2}{2}-\frac{(q_2^y)^2}{2}-d_2q_2^xq_2^y}=\sum_n \lambda_{2,n}\psi_n(q^x_2)\psi_n(q^y_2)$$
$$\vdots$$
$$e^{-\frac{(q_m^x)^2}{2}-\frac{(q_m^y)^2}{2}-d_mq_m^xq_m^y}=\sum_n \lambda_{m,n}\psi_n(q^x_m)\psi_n(q^y_m)$$
where
$\lambda_{i,n}=(\frac{2}{\nu_i+1})^{1/2}(\frac{\nu_i-1}{\nu_i+1})^{n/2}$.

Therefore the entropy of each part can be written
$$S(\rho_i)=\frac{\nu_i +1}{2} log(\frac{\nu_i +1}{2})-\frac{\nu_i
-1}{2}log(\frac{\nu_i -1}{2})$$
\begin{equation}
=\frac{(\frac{1}{1-d_i^2})^{1/2} +1}{2}
log(\frac{(\frac{1}{1-d_i^2})^{1/2}
+1}{2})-\frac{(\frac{1}{1-d_i^2})^{1/2}
-1}{2}log(\frac{(\frac{1}{1-d_i^2})^{1/2} -1}{2})
\end{equation}
So the total entropy is
\begin{equation}
S(\rho)=\sum_i S(\rho_i)
\end{equation}

\section{Entropy of entanglement by using generalized Schur complement method}

\setlength{\unitlength}{0.75cm}
\begin{picture}(6,6)
\linethickness{0.075mm}  \put(8,5.5){$m$} \put(11,5.5){$n$}
 \put(8,4){\circle*{0.2}}
\put(8,4.5){\circle*{0.2}}
\put(8,1.5){\circle*{0.2}}\put(8,2){\circle*{0.2}}\put(8,2.5){\circle*{0.2}}
\put(11, 4){\circle*{0.2}} \put(11,4.5){\circle*{0.2}} \put(11,
5){\circle*{0.2}} \put(11,3){\circle*{0.2}} \put(11,
2.5){\circle*{0.2}} \put(11,2){\circle*{0.2}} \put(11,
1.5){\circle*{0.2}} \put(8,4){\line(1, 0){3}}
\put(8,4.5){\line(1,0){3}} \put(8,4.5){\line(6,1){3}}
\put(8,4){\line(3,1){3.1}} \put(11,4){\line(0,1){0.5}}
\put(11,2){\line(0,1){0.5}} \put(8,2){\line(0,1){0.5}}
\put(11,4){\oval(2,2)[r]} \put(11,3){\oval(2,3)[r]}
\put(8,3){\oval(2,3)[l]} \put(11,2.5){\oval(1,1)[r]}
\put(11,2){\oval(1,1)[r]} \put(8,2){\oval(1,1)[l]}
\put(8,3){\oval(2,2)[l]}
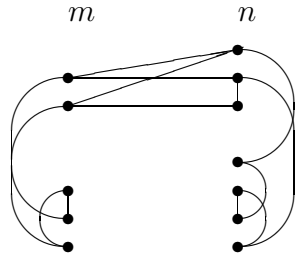
\put(3,0){\footnotesize FIG I: An example of graph partitioning in a graph}

\end{picture}

We want to calculate the entanglement entropy between two parts in an arbitrary graph. Suppose there are $m$ ($n$) nodes in the first (second) part. There are $m_1$ ($n_1$) nodes in the first (second) part which are not connected to the nodes of other part. one can separate the nodes of each part $m,n$ into two subsets, so there are four subsets which have
$m_1,m_2,n_2,n_1$ nodes, respectively. An example of this partitioning is shown in figure 1 and 2.

\setlength{\unitlength}{0.75cm}
\begin{picture}(6,4)
\linethickness{0.075mm} \put(5,1){\oval(1.8,3)}
\put(8,1){\oval(1.8,3)} \put(11,1){\oval(1.8,3)}
\put(14,1){\oval(1.8,3)} \put(5,3){$m_1$} \put(8,3){$m_2$}
\put(11,3){$n_2$} \put(14,3){$n_1$} \put(5, 0){\circle*{0.2}}
\put(5,1){\circle*{0.2}} \put(5, 2){\circle*{0.2}}
\put(8,0){\circle*{0.2}} \put(8,1){\circle*{0.2}}  \put(11,
0){\circle*{0.2}} \put(11,1){\circle*{0.2}} \put(11,
2){\circle*{0.2}} \put(14,0){\circle*{0.2}} \put(14,
0.5){\circle*{0.2}} \put(14,1){\circle*{0.2}} \put(14,
1.5){\circle*{0.2}} \put(5,1){\line(1,0){3}}
\put(5,1){\line(0,1){1}} \put(5,0){\line(1,0){3}}
 \put(8,1){\line(1,0){3}}
\put(8,0){\line(1,0){3}} \put(8,0){\line(3,2){3.1}}
\put(8,1){\line(3,1){3.1}} \put(11,0){\line(0,1){1}}
\put(14,0.5){\line(0,1){0.5}} \put(11,2){\line(6,-1){3}}
\put(11,1){\line(3,-1){3.1}} \put(5,1){\oval(1.5,2)[l]}
\put(14,0.5){\oval(1,1)[r]} \put(14,1){\oval(1,1)[r]}
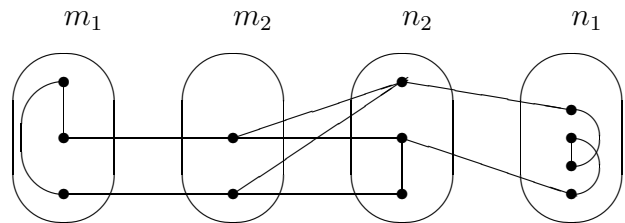
\put(0,-1){\footnotesize FIG II: The graph in FIG I is separated to four subsets. $m_1$ nodes in the part $(m_1,m_2)$ are not connected}
\put(0,-2){\footnotesize to the nodes of part $(n_1,n_2)$. Also the $n_1$ nodes in part $(n_1,n_2)$ are not connected to the nodes of part $(m_1,m_2)$.}

\end{picture}
\newline
\newline
\newline
The potential matrix of the system is:

\begin{equation}
V=I+2gL=\left(\begin{array}{cccc}
          V_{11}& V_{12} & 0 &0\\
            V_{12}^T & V_{22} & V_{23} & 0 \\
            0 & V_{23}^T & V_{33} & V_{34}\\
            0 & 0 & V_{34}^T & V_{44} \\
          \end{array}\right)
\end{equation}

So the wave function will be
\begin{equation}
\psi(x,y,z,w)=A_g exp(-\frac{1}{2}\left(\begin{array}{cccc}x & y & z & w \\
\end{array}\right)\left(\begin{array}{cccc}
          V_{11}& V_{12} & 0 &0\\
            V_{12}^T & V_{22} & V_{23} & 0 \\
            0 & V_{23}^T & V_{33} & V_{34}\\
            0 & 0 & V_{34}^T & V_{44} \\
          \end{array}\right)\left(\begin{array}{c}
          x\\
          y\\
          z\\
          w\\
\end{array}\right))
\end{equation}

Then by using the Generalized Schur complement theorem, we can
write

$$\left(\begin{array}{cccc}
          V_{11}& V_{12} & 0 &0\\
            V_{12}^T & V_{22} & V_{23} & 0 \\
            0 & V_{23}^T & V_{33} & V_{34}\\
            0 & 0 & V_{34}^T & V_{44} \\
          \end{array}\right)=$$
\begin{equation}
          =\left(\begin{array}{cccc}
          I_{m_1}& 0 & 0 & 0\\
            V_{12}^TV_{11}^{-1}& I_{m_2} & 0 & 0 \\
           0 & 0 & I_{n_2}& V_{34}V_{44}^{-1}\\
           0 & 0 & 0 & I_{n_1}\\
          \end{array}\right)\left(\begin{array}{cccc}
          V_{11}& 0 & 0 & 0 \\
           0 & \widetilde{V}_{22} & V_{23} & 0\\
            0 & V_{23}^T & \widetilde{V}_{33}  & 0\\
            0 & 0 & 0 & V_{44} \\
          \end{array}\right)\left(\begin{array}{cccc}
          I_{m_1}& V_{11}^{-1}V_{12} & 0 & 0 \\
            0& I_{m_2}& 0 & 0 \\
            0 & 0 & I_{n_2} & 0 \\
            0 & 0 & V_{44}^{-1}V_{34}^T & I_{n_1} \\
          \end{array}\right)
\end{equation}
Where $\widetilde{V}_{22}=V_{22}-V_{12}^T V_{11}^{-1}V_{12}$ and
$\widetilde{V}_{33}=V_{33}-V_{34}V_{44}^{-1}V_{34}^T$.

It can be shown that by this transformation, the Schmidt numbers
and the entropy will be unchanged, because the coordinates are
transformed to
$$x'=x+V_{11}^{-1}V_{12}y$$
$$y'=y$$
$$z'=z$$
\begin{equation}
w'=w+V_{44}^{-1}V_{34}^T z
\end{equation}

So the wave function will be
\begin{equation}
\psi(x,y,z,w)=A_g exp(-\frac{1}{2}\left(\begin{array}{cc} y & z \\
\end{array}\right)\left(\begin{array}{cc}
          \widetilde{V}_{22} & V_{23}\\
          V_{23}^T & \widetilde{V}_{33}\\
          \end{array}\right)\left(\begin{array}{c}
          y\\
          z\\
\end{array}\right)+(x+Fy)^TV_{11}(x+Fy)+(w+Gz)^TV_{44}(w+Gz))
\end{equation}
Where $F=V_{11}^{-1}V_{12}$ and $G=V_{44}^{-1}V_{34}^T$. So after
using the equation (3-13) it can be written as
\begin{equation}
(1-t^2)^{1/2}\sum t^n
\psi_n(y)\psi_n(z)e^{-\frac{1}{2}(x+Fy)^TV_{11}(x+Fy)}e^{-\frac{1}{2}(w+Gz)^TV_{44}(w+Gz)}
\end{equation}
The new states will be
$$\psi_n(y,x)=\psi_n(y)e^{-\frac{1}{2}(x+Fy)^TV_{11}(x+Fy)}$$
\begin{equation}
\psi_n(z,w)=\psi_n(z)e^{-\frac{1}{2}(w+Gz)^TA_{44}(w+Gz)}
\end{equation}
The new states are orthogonal, because
$$\int e^{-\frac{1}{2}(x+Fy)^TV_{11}(x+Fy)} dx=\int e^{-\frac{1}{2} \eta^TV_{11}\eta d\eta}=\sqrt{\frac{\pi^{m_1}}{\prod_{i=1}^{m_1}\lambda_i^{11}}}$$

Also
\begin{equation}
\int e^{-\frac{1}{2}(w+Gz)^TV_{44}(w+Gz)}
dw=\sqrt{\frac{\pi^{n_1}}{\prod_{i=1}^{n_1}\lambda_i^{44}}}
\end{equation}

\subsection{Theorem I}
Suppose all  graphs  which their
blocks $V_{12}$, $V_{23}$ and $V_{34}$ are complete graphs. All of these graphs have the same Schmidt number and the same entanglement entropy.

\textit{proof}: According to the supposition of the theorem, the
blocks $V_{12}$, $V_{23}$ and $V_{34}$ are complete graphs. then
the potential matrix will be

\begin{equation}
\left(\begin{array}{cccc}
          V_{11}& J_{m_1\times m_2} & 0 &0\\
            J_{m_2 \times m_1} & V_{22} & J_{m_2\times n_2} & 0 \\
            0 & J_{n_2 \times m_2} & V_{33} & J_{n_2 \times n_1}\\
            0 & 0 & J_{n_1 \times n_2} & V_{44} \\
          \end{array}\right)
\end{equation}

By applying the generalized Schur complement

$$\left(\begin{array}{cccc}
          V_{11}& V_{12} & 0 &0\\
            V_{12}^T & V_{22} & V_{23} & 0 \\
            0 & V_{23}^T & V_{33} & V_{34}\\
            0 & 0 & V_{34}^T & V_{44} \\
          \end{array}\right)=$$
\begin{equation}
          =\left(\begin{array}{cccc}
          I_{m_1}& 0 & 0 & 0\\
            V_{12}^TV_{11}^{-1}& I_{m_2} & 0 & 0 \\
           0 & 0 & I_{n_2}& V_{34}V_{44}^{-1}\\
           0 & 0 & 0 & I_{n_1}\\
          \end{array}\right)\left(\begin{array}{cccc}
          V_{11}& 0 & 0 & 0 \\
           0 & \widetilde{V}_{22} & V_{23} & 0\\
            0 & V_{23}^T & \widetilde{V}_{33}  & 0\\
            0 & 0 & 0 & V_{44} \\
          \end{array}\right)\left(\begin{array}{cccc}
          I_{m_1}& V_{11}^{-1}V_{12} & 0 & 0 \\
            0& I_{m_2}& 0 & 0 \\
            0 & 0 & I_{n_2} & 0 \\
            0 & 0 & V_{44}^{-1}V_{34}^T & I_{n_1} \\
          \end{array}\right)
\end{equation}
Where $\widetilde{V}_{22}=V_{22}-V_{12}^T V_{11}^{-1}V_{12}$ and
$\widetilde{V}_{33}=V_{33}-V_{34}V_{44}^{-1}V_{34}^T$.

These kind of graphs after applying the above transformations will
be

\begin{equation}
\left(\begin{array}{cccc}
          V_{11}& 0 & 0 &0\\
            0 & V_{22}-J^TV_{11}^{-1}J & J_{m_2\times n_2} & 0 \\
            0 & J_{n_2 \times m_2} & V_{33}-JV_{44}^{-1}J^T & 0\\
            0 & 0 & 0 & V_{44} \\
          \end{array}\right)
\end{equation}

The left eigenvectors of $J_{m_2\times n_2}$ are the eigenvectors
of $\widetilde{A}_{22}$ and the right eigenvectors of
$J_{m_2\times n_2}$ are the eigenvectors of $\widetilde{V}_{33}$,
so if we rewrite the potential matrix in the basis of eigenvectors
of $J$, then just the first element of matrices
$\widetilde{V}_{22}$ and $\widetilde{V}_{33}$ are effective in the
entropy.

Then the first element of
$\widetilde{V}_{22}=V_{22}-J^TV_{11}^{-1}J$ will be the sum of the
first element of $V_{22}$ and $-J_{m_1\times
m_2}^TV_{11}^{-1}J_{m_1\times m_2}$, but the left eigenvectors of
$J_{m_1\times m_2}$ are the eigenvectors of $V_{11}$, So only the
first element of $V_{11}$ is in the $-J_{m_1\times
m_2}^TV_{11}^{-1}J_{m_1\times m_2}$. These arguments are true for
the $\widetilde{V}_{33}=V_{33}-JV_{44}^{-1}J^T$.

So the parameter $d$ will be

\begin{equation}
d=\frac{2g\sqrt{m_2n_2}}{\sqrt{1+2g(m_1+n_2)-\frac{4g^2m_1m_2}{1+2gm_2}}\sqrt{1+2g(m_2+n_1)-\frac{4g^2n_1n_2}{1+2gn_2}}}
\end{equation}

\textbf{\textit{corollary I}}
Let in a graph the parameters $m_1=n_1=0$ and $m_2=m, n_2=n$, So the graph has only two subsets. From theorem I, all of these kinds of graphs with the complete connections between two parts, have the same Schmidt number and the same entanglement entropy. Their parameter $d$ are

\begin{equation}
d=\frac{2g\sqrt{mn}}{\sqrt{(1+2gm)(1+2gn)}}
\end{equation}

\textbf{\textit{corollary II}}
Suppose two  graphs with the same numbers $m_1,m_2,n_1,n_2$  and in their adjacency matrices, the blocks $V_{12}$ are  complete graphs ( i.e. they are $J_{m_1\times m_2}$ ), and the other blocks except $V_{11}$ are the same. So their bipartite entanglement entropies will be identical.
This corollary is satisfied, for the case that in two graphs, the blocks $V_{34}$ are complete graph and the blocks $V_{44}$ are different, and the other blocks are the same.

\subsection{examples}
\textit{Example I: Complete Graph}\\In this example, we consider
the N-vertex complete graph $K_N$, then we separate it into two parts with the vertices
$m$ , $N-m$. we can calculate the entanglement entropy and Schmidt
number by using corollary I. The parameter $d$ is

\begin{equation}
d_{m,N-m}=\frac{2g\sqrt{m(N-m)}}{\sqrt{1+2mg}\sqrt{1+2g(N-m)}}
\end{equation}

\textit{Example II: Path-Graph} We consider the
entanglement entropy in the path graph with even $N$ vertices. We choose
two equal parts from it with $n=\frac{N}{2}$ vertices. So, the blocks
of potential matrix, according to (3-5) are:
\begin{equation}
A=\left(%
\begin{array}{ccccc}
  1+2g & -2g & 0 & ... & 0 \\
  -2g & 1+4g & -2g & ... & 0 \\
  0 & -2g & 1+4g & ... & 0 \\
  \vdots & \vdots & \vdots & \ddots & \vdots \\
  0 & ... & 0 & -2g & 1+4g \\
\end{array}%
\right), \quad
B=\left(%
\begin{array}{cccc}
  0 & 0 & ... & 0 \\
  0 & 0 & ... & 0 \\
  \vdots & \vdots & \ddots & \vdots \\
  -2g & 0 & ... & 0 \\
\end{array}%
\right)
\end{equation}
$$C=\left(%
\begin{array}{ccccc}
  1+4g & -2g & 0 & ... & 0 \\
  -2g & 1+4g & -2g & ... & 0 \\
  0 & -2g & 1+4g & ... & 0 \\
  \vdots & \vdots & \vdots & \ddots & \vdots \\
  0 & ... & 0 & -2g & 1+2g \\
\end{array}%
\right)$$ By applying $n-1$ generalized schur complement steps
from (4-24), the schmidt number can be calculated
\begin{equation}
d=\frac{2g}{1+4g-\frac{4g^{2}}{1+4g-\frac{4g^{2}}{1+4g-\frac{4g^{^{2}}}{1+4g-...\frac{4g^{2}}{1+2g}}}}}=\frac{Q_{n-1}(x)}{Q_{n}(x)}
\end{equation}
The parameter $x$ defined as $x=2+\frac{1}{2g}$ And $Q_{n}$
defined recurrently by
$$Q_{0}(x)=1\quad,\quad Q_{1}(x)=x-1$$
\begin{equation}
Q_{n}(x)=xQ_{n-1}(x)-Q_{n-2}(x)
\end{equation}
for $n\geq2$. Where above polynomial can be shown that is the
second type Tchebichef polynomials as following
\begin{equation}
Q_{n}(\theta)=\frac{\sin(n+1)\theta}{\sin\theta}
\end{equation}
By replacing in equation $(4-36)$, we can obtain
\begin{equation}
x=2\cos\theta
\end{equation}
So,
\begin{equation}
Q_{n}(\theta)=\frac{\sin(n+1)\cos^{-1}\frac{x}{2}}{\sin\cos^{-1}\frac{x}{2}}
\end{equation}
By definition $Q_{n}=U_{n}(\frac{x}{2})$ as Tchebichef
polynomials, the Schmidt number can be written as following
\begin{equation}
d_{n}=\frac{U_{n-1}(1+\frac{1}{4g})}{U_{n}(1+\frac{1}{4g})}
\end{equation}

\textit{Example III: Barbell graph}:

The n-Barbell graph is the simple graph obtained by connecting two copies of complete graph $K_n$ by a bridge. The generalized Barbell graph can be obtained by connecting two complete graph $K_{l_1}$ and $K_{l_2}$ by a bridge. In this figure, two complete graphs $K_3$ and $K_4$ are coalesced. This graph is an example which satisfies the theorem I.

\setlength{\unitlength}{0.75cm}
\begin{picture}(6,4)
\linethickness{0.075mm} \put(5,1){\oval(1.8,3)}
\put(8,1){\oval(1.8,3)} \put(11,1){\oval(1.8,3)}
\put(14,1){\oval(1.8,3)} \put(5,3){$m_1$} \put(8,3){$m_2$}
\put(11,3){$n_2$} \put(14,3){$n_1$} \put(5, 0){\circle*{0.2}}
\put(5, 2){\circle*{0.2}}
 \put(8,1){\circle*{0.2}}   \put(11,1){\circle*{0.2}}  \put(14,0){\circle*{0.2}} \put(14,
2){\circle*{0.2}} \put(14,1){\circle*{0.2}}
 \put(5,0){\line(0,1){2}}
\put(5,2){\line(3,-1){3}} \put(5,0){\line(3,1){3}}
 \put(8,1){\line(1,0){3}}
\put(11,1){\line(1,0){3}} \put(11,1){\line(3,1){3.1}}
\put(11,1){\line(3,-1){3.1}}

\put(14,0){\line(0,1){1}} \put(14,1){\line(0,1){1}}

\put(14,1){\oval(1.5,2)[r]}
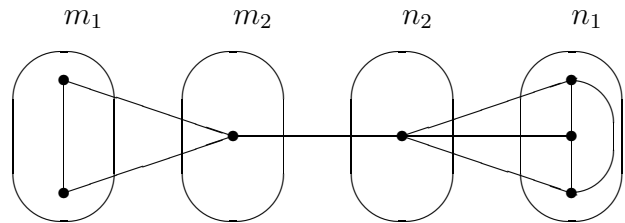
\put(2,-1){\footnotesize FIG III: A generalized Barbell graph which is compound of $K_3$ and $K_4$.}
\end{picture}
\newline
\newline
The graph in FIG IV, is another example
of graphs which satisfies the Theorem I. It is the coalescence of
two star graphs. In the following figure two star graphs ($S_3$ and $S_4$) are
coalesced.

\setlength{\unitlength}{0.75cm}
\begin{picture}(6,4)
\linethickness{0.075mm} \put(5,1){\oval(1.8,3)}
\put(8,1){\oval(1.8,3)} \put(11,1){\oval(1.8,3)}
\put(14,1){\oval(1.8,3)} \put(5,3){$m_1$} \put(8,3){$m_2$}
\put(11,3){$n_2$} \put(14,3){$n_1$} \put(5, 0){\circle*{0.2}}
\put(5, 2){\circle*{0.2}}
 \put(8,1){\circle*{0.2}}   \put(11,1){\circle*{0.2}}  \put(14,0){\circle*{0.2}}  \put(14,1){\circle*{0.2}} \put(14,2){\circle*{0.2}}

\put(5,2){\line(3,-1){3}} \put(5,0){\line(3,1){3}}
 \put(8,1){\line(1,0){3}}
\put(11,1){\line(1,0){3}} \put(11,1){\line(3,1){3.1}}
\put(11,1){\line(3,-1){3.1}}
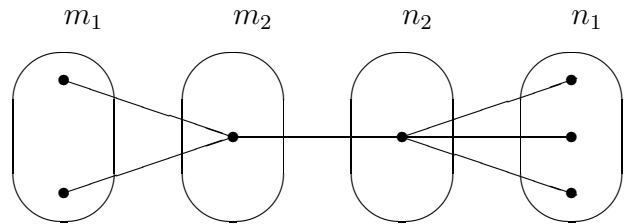
\put(1,-1){\footnotesize FIG IV: A graph which is compound of $S_3$ and $S_4$. This graph satisfies theorem I.}
\end{picture}
\newline
\newline
\textit{example IV: lollipop graph}

An example for corollary II, is \textit{lollipop graph}. The
$(m,n)$ lollipop graph is the graph obtained by joining a complete
graph $K_m$ to a path graph $P_n$ with a bridge. In this figure, a
complete graphs $K_5$ and path graph $P_4$ are coalesced.

\setlength{\unitlength}{0.75cm}
\begin{picture}(6,4)
\linethickness{0.075mm} \put(5.5,1){\oval(1.5,3)}
\put(7,1){\oval(1.2,3)} \put(9,1){\oval(1.8,3)}
\put(12,1){\oval(4,3)} \put(6,3){$m_1$} \put(7,3){$m_2$}
\put(9,3){$n_2$} \put(12,3){$n_1$} \put(5,0.5){\circle*{0.2}}
\put(5,1.5){\circle*{0.2}}
 \put(6,0){\circle*{0.2}}
 \put(6,2){\circle*{0.2}}
 \put(7,1){\circle*{0.2}}
 \put(9,1){\circle*{0.2}}
 \put(11,1){\circle*{0.2}}
 \put(13,1){\circle*{0.2}}
\put(5,0.5){\line(0,1){1}} \put(7,1){\line(1,0){3}}
\put(9,1){\line(1,0){3}} \put(11,1){\line(1,0){2}}
\put(5,0.5){\line(4,1){2}} \put(5,1.5){\line(4,-1){2}}
\put(5,0.5){\line(2,-1){1}} \put(5,1.5){\line(2,1){1}}

\put(6,0){\line(1,1){1}} \put(6,2){\line(1,-1){1}}

\put(6,0){\line(0,1){2}}

\put(5,0.5){\line(2,3){1}} \put(5,1.5){\line(2,-3){1}}
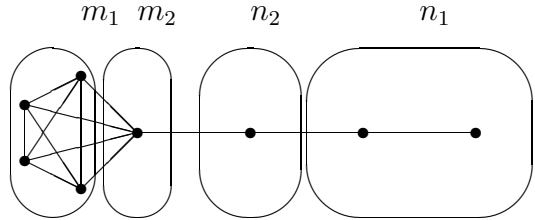
\put(3,-1){\footnotesize FIG V: The Lollipop graph which is compound of $K_5$ and $P_4$.}
\end{picture}
\newline
\newline
The adjacency matrix of a lollipop(m,n) is in the form
\begin{equation}
A=\left(\begin{array}{cccc}
          J_{m-1\times m-1}& J_{m-1 \times 1} & 0 &0\\
            J_{1\times m-1} & 0 & 1 & 0 \\
            0 & 1 & 0 & 1\\
            0 & 0 & 1 & A_{p(n-1)} \\
          \end{array}\right)
\end{equation}

So the potential matrix will be
\begin{equation}
V=\left(\begin{array}{cccc}
         (1+2gm)I-2gJ& -2gJ_{m-1 \times 1} & 0 &0\\
            -2gJ_{1\times m-1} & 1+2gm & -2g & 0 \\
            0 & -2g & 1+2g(2) & -2g\\
            0 & 0 & -2g & p(n-1) \\
          \end{array}\right)
\end{equation}

Where $p(n)$ is as following.

\begin{equation}
p(n)=\left(\begin{array}{ccccc}
         1+4g& -2g & 0 & \ldots & 0\\
       -2g & 1+4g & -2g & \ldots & 0 \\
            0 & -2g & 1+4g & \ddots & 0 \\
            0& \vdots & \ddots & \ddots &-2g \\
            0 & 0 & \ldots & -2g & 1+2g \\
          \end{array}\right)
\end{equation}

Again by using Schur complement method for this graph, the
potential matrix is transformed to
\begin{equation}
V=\left(\begin{array}{cccc}
         (1+2gm)I_{m-1\times m-1}-2gJ_{m-1\times m-1}& 0 & 0 &0\\
            0 & \widetilde{V_{22}} & -2g & 0 \\
            0 & -2g & C(n) & 0\\
            0 & 0 & 0 & D(n-1) \\
          \end{array}\right)
\end{equation}
Where
\begin{equation}
D(n-1)=\left(\begin{array}{cccc}
         C(n-1)& 0 &  \ldots & 0\\
       0 & \ddots &  \ldots & 0 \\
            0& \ldots & C(2) & 0 \\
            0 & 0 & \ldots & C(1) \\
          \end{array}\right)
\end{equation}
And
\begin{equation}
C(n)=1+4g-\frac{4g^{2}}{1+4g-\frac{4g^{2}}{1+4g-\frac{4g^{^{2}}}{1+4g-...\frac{4g^{2}}{1+2g}}}}
\end{equation}
from (4.41), $C(n)$ is
\begin{equation}
C(n)=2g\frac{U_n(1+\frac{1}{4g})}{U_{n-1}(1+\frac{1}{4g})}
\end{equation}
And
\begin{equation}
\widetilde{V_{22}}=1+2gm-\frac{4g^2(1+2g)(m-1)}{(1+2mg)(1+2g)}-\frac{4g^2(2g)(m-1)^2}{(1+2mg)(1+2g)}
\end{equation}

So the parameter $d$ will be
\begin{equation}
d=\frac{2g}{\sqrt{C(n-1)}\sqrt{\widetilde{V_{22}}}}
\end{equation}

The graph in FIG VI will be obtained by joining a star graph $S_m$
to a path graph $P_n$ with a bridge. It's entanglement entropy is equal to lollipop graph (compound of $K_m$ and $P_n$) from the corollary II.

\setlength{\unitlength}{0.75cm}
\begin{picture}(6,4)
\linethickness{0.075mm} \put(5.5,1){\oval(1.5,3)}
\put(7,1){\oval(1.2,3)} \put(9,1){\oval(1.8,3)}
\put(12,1){\oval(4,3)} \put(6,3){$m_1$} \put(7,3){$m_2$}
\put(9,3){$n_2$} \put(12,3){$n_1$} \put(5,0.5){\circle*{0.2}}
\put(5,1.5){\circle*{0.2}}
 \put(6,0){\circle*{0.2}}
 \put(6,2){\circle*{0.2}}
 \put(7,1){\circle*{0.2}}
 \put(9,1){\circle*{0.2}}
 \put(11,1){\circle*{0.2}}
 \put(13,1){\circle*{0.2}}

\put(7,1){\line(1,0){2}} \put(9,1){\line(1,0){2}}
\put(11,1){\line(1,0){2}}

\put(5,0.5){\line(4,1){2}} \put(5,1.5){\line(4,-1){2}}

\put(6,0){\line(1,1){1}} \put(6,2){\line(1,-1){1}}
\put(2,-1){\footnotesize FIG VI: A graph which is compound of $S_5$ and $P_4$.}
\end{picture}
\newline
\newline
\section{Conductance and bipartite entanglement}
By considering the all partitions of graph into two parts, suppose
we have $m$ vertices in the smaller part and $N-m$ vertices in the
other part. Then the conductance is
\begin{equation}
\alpha(G)=\min{\frac{|E(m,n)|}{m}}_{m\leq\frac{N}{2}}
\end{equation}

Where $|E(m,n)|$ is the number of connections (edges) between two
parts.

\subsection{Example I: Complete graph}
\begin{equation}
\alpha(G)=\min{\frac{|E(m,n)|}{m}}_{m\leq\frac{N}{2}}=\min{\frac{m(N-m)}{m}}_{m\leq\frac{N}{2}}=\min{N-m}_{m\leq\frac{N}{2}}=\{\begin{array}{cc}
  \frac{N}{2} & even N \\
  \frac{N+1}{2} & odd N \\
\end{array}
\end{equation}

This partition gives the maximum entropy of entanglement in
complete graph.
\\By calculating entanglement entropy, we have
The Schmidt number:
$$d_{n=1,N-1}=\frac{2g\sqrt{N-1}}{\sqrt{1+2g}\sqrt{1+2g(N-1)}}$$
$$d_{n=2,N-2}=\frac{2g\sqrt{2(N-2)}}{\sqrt{1+4g}\sqrt{1+2g(N-2)}}$$
$$\vdots$$
\begin{equation}
d_{n=m,N-m}=\frac{2g\sqrt{m(N-m)}}{\sqrt{1+2mg}\sqrt{1+2g(N-m)}}
\end{equation}
After calculating entropy, we have \\\textit{N=Odd}
\begin{equation}
S_{n=1,N-1}<S_{n=2,N-2}<...<S_{n=\frac{N-1}{2},\frac{N+1}{2}}
\end{equation}
 \\\textit{N=Even}
\begin{equation}
S_{n=1}=S_{n=N-1}<S_{n=2}=S_{n=N-2}<...<S_{n=\frac{N}{2}}
\end{equation}
In complete graph, the partition that has maximum entanglement
entropy, has minimum conductance number.

\subsection{Example II: Path  graph}
In the path graph (with even number of vertices), the conductance
will be achieved when half of neighbor vertices are in one part
and the other neighbor vertices are in the other part.

\begin{equation}
\alpha(G)=\frac{1}{\frac{N}{2}}=\frac{2}{N}
\end{equation}
\setlength{\unitlength}{0.75cm}
\begin{picture}(6,4)
\linethickness{0.075mm}

\put(9,0){\circle*{0.2}} \put(9,1){\circle*{0.2}}
\put(9,2){\circle*{0.2}} \put(9,3){\circle*{0.2}}

\put(11,0){\circle*{0.2}} \put(11,1){\circle*{0.2}}
\put(11,2){\circle*{0.2}} \put(11,3){\circle*{0.2}}

\put(9,0){\line(1,0){2}} \put(9,0){\line(0,1){1}}
\put(11,0){\line(0,1){1}} \put(9,2){\line(0,1){1}}
\put(11,2){\line(0,1){1}}

\put(9,1.5){\oval(1.8,4)} \put(11,1.5){\oval(1.8,4)}

\put(9,1.2){$\vdots$} \put(11,1.2){$\vdots$}
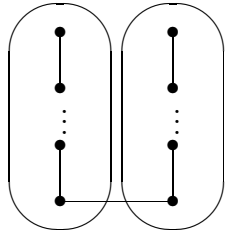
\put(0,-1){\footnotesize FIG VII: The Path graph with even number of vertices. In this partitioning the half}
\put(0,-2){\footnotesize  of alternative vertices are in one part, and the rest of vertices are in the other part.}
\end{picture}
\newline
\newline
\newline
It can be shown that this partition gives the minimum entropy of
entanglement with respect to all kinds of partitions of path graph.

\subsection{Example III: Star graph}
In the general star graphs with $N$ vertices, different partitions give the conductance:
The number of $i$ vertices ($i=1,2,\ldots,\frac{N}{2}$ for even $N$ and $\frac{N-1}{2}$ for odd $N$), are in one part and the central vertex and
the other vertices are in other part.

The conductance of graph is
\begin{equation}
\alpha(G)=1
\end{equation}
All of these partitions satisfy the theorem I. Suppose $i$ vertices are in one part, and the central vertex and the other vertices are in the other part. So $m_1=0,m_2=i,n_2=1,n_1=N-i-1$ and the parameter $d$ from (4-33) is
\begin{equation}
d=\frac{2g\sqrt{i}}{\sqrt{(1+2g)(1+2g(N-1))-4g^2(N-i-1)}}
\end{equation}

So the parameter $\alpha(G)$ is the same for $\frac{N}{2}$ kinds of partitions,
but the entanglement entropy between two parts, are different for
all of these partitions.

\setlength{\unitlength}{0.75cm}
\begin{picture}(6,5)
\linethickness{0.075mm}
\put(0.5,4){$S_4(1)$}
\put(6.5,4){$S_4(2)$}
\put(12.5,4){$S_4(3)$}

\put(0,2){\circle*{0.2}} \put(2,1){\circle*{0.2}}
\put(2,2){\circle*{0.2}} \put(2,3){\circle*{0.2}}

\put(6,2){\circle*{0.2}} \put(6,3){\circle*{0.2}}
\put(8,2){\circle*{0.2}} \put(8,3){\circle*{0.2}}

\put(12,2){\circle*{0.2}} \put(14,1){\circle*{0.2}}
\put(14,2){\circle*{0.2}} \put(14,3){\circle*{0.2}}

\put(0,2){\line(1,0){2}} \put(0,2){\line(2,1){2}}
\put(0,2){\line(2,-1){2}} \put(6,3){\line(0,-1){1}}
\put(6,3){\line(1,0){2}} \put(6,3){\line(2,-1){2}}
\put(12,2){\line(1,0){2}} \put(14,1){\line(0,1){2}}

\put(0,2){\oval(1.8,3)} \put(2,2){\oval(1.8,3)}
\put(6,2.5){\oval(1.8,2)} \put(8,2.5){\oval(1.8,2)}
\put(12,2){\oval(1.8,3)} \put(14,2){\oval(1.8,3)}

\put(6,0){$\alpha(G)=1$} \put(12,0){$\alpha(G)=1$}

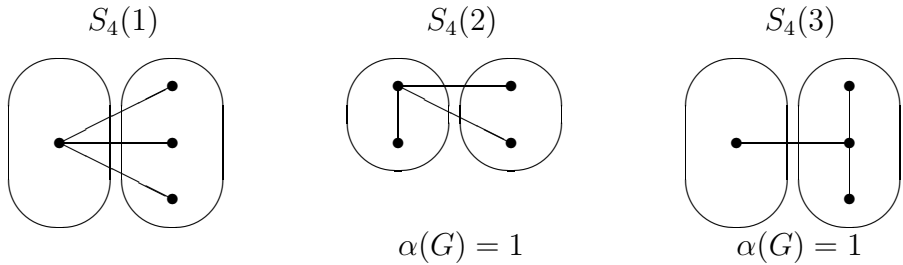
\put(0,-1){\footnotesize  FIG VIII: All kinds of partitioning in $S_4$ graph. Two different partition gives the conductance,}
\put(0,-2){\footnotesize  But their entanglement entropies are not the same. $S(\rho_{S_4(1)})>S(\rho_{S_4(2)})>S(\rho_{S_4(3)})$}
\end{picture}
\newline
\newline
\newline
\subsection{Example IV: kite graph}
In the Kite graph, the conductance will be $\alpha(G)=\frac{3}{2}$
when two neighbor vertices are in one part and the other  two
neighbor vertices are in the other part. The entanglement entropy
for this partition is not maximum, also it is not minimum.

\setlength{\unitlength}{0.75cm}
\begin{picture}(6,5)
\linethickness{0.075mm}
\put(0.5,4){$Kite(1)$}
\put(6.5,4){$Kite(2)$}
\put(12.5,4){$Kite(3)$}
\put(18.5,4){$Kite(4)$}
\put(0,2){\circle*{0.2}} \put(0,3){\circle*{0.2}}
\put(2,2){\circle*{0.2}} \put(2,3){\circle*{0.2}}

\put(6,2){\circle*{0.2}} \put(6,3){\circle*{0.2}}
\put(8,2){\circle*{0.2}} \put(8,3){\circle*{0.2}}

\put(12,2){\circle*{0.2}} \put(14,1){\circle*{0.2}}
\put(14,2){\circle*{0.2}} \put(14,3){\circle*{0.2}}

\put(18,2){\circle*{0.2}} \put(20,1){\circle*{0.2}}
\put(20,2){\circle*{0.2}} \put(20,3){\circle*{0.2}}

\put(0,2){\line(1,0){2}} \put(0,2){\line(0,1){1}}
\put(0,3){\line(1,0){2}} \put(0,2){\line(2,1){2}}
\put(0,3){\line(2,-1){2}}

\put(6,2){\line(1,0){2}} \put(6,2){\line(0,1){1}}
\put(6,3){\line(1,0){2}} \put(6,3){\line(2,-1){2}}
\put(8,2){\line(0,1){1}}

\put(12,2){\line(1,0){2}} \put(12,2){\line(2,1){2}}
\put(12,2){\line(2,-1){2}} \put(14,1){\line(0,1){2}}

\put(18,2){\line(2,-1){2}} \put(18,2){\line(2,1){2}}
\put(20,1){\line(0,1){2}} \put(20,2){\oval(0.5,2)[r]}

\put(0,2.5){\oval(1.8,2)} \put(2,2.5){\oval(1.8,2)}
\put(6,2.5){\oval(1.8,2)} \put(8,2.5){\oval(1.8,2)}
\put(12,2){\oval(1.8,3)} \put(14,2){\oval(1.8,3)}
\put(18,2){\oval(1.8,3)} \put(20,2){\oval(1.8,3)}

\put(6,0){$\alpha(G)=\frac{3}{2}$}
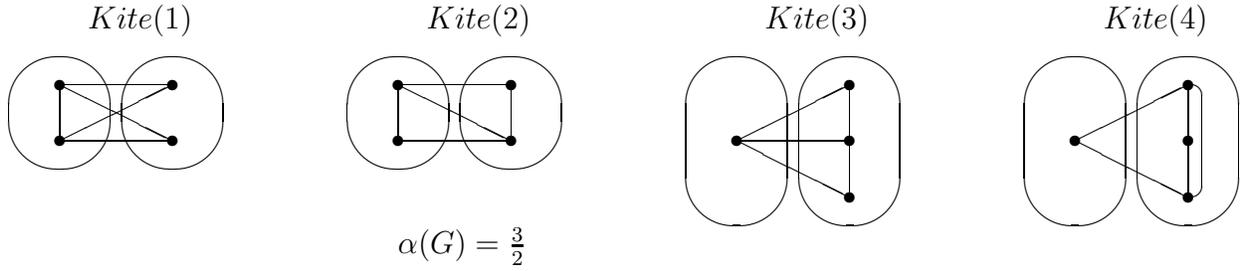
\put(0,-1){\footnotesize  FIG IX: All kinds of partitioning in Kite graph. The conductance is for partitioning $Kite(2)$}
\put(0,-2){\footnotesize  The entanglement entropies are $S(\rho_{Kite(1)})>S(\rho_{Kite(2)})>S(\rho_{Kite(3)})>S(\rho_{Kite(3)})$}
\end{picture}
\newline
\newline
\newline
\subsection{Example V: Square graph}
In the Square graph, the conductance will be $\alpha(G)=1$ when
two neighbor vertices are in one part and the other  two neighbor
vertices are in the other part.

The entanglement entropy for Square graph is maximum when two
vertices in the diameter of square are in a part, So it is
bipartite complete graph with number of vertices: $m=2,n=2$. The
entanglement entropy for Square graph is minimum when one of
vertices is in one part and the other vertices are in the other
part.

The entanglement entropy for partition which gives the
conductance, is not maximum or minimum.

\setlength{\unitlength}{0.75cm}
\begin{picture}(6,5)
\linethickness{0.075mm}
\put(0.5,4){$Square(1)$}
\put(6.5,4){$Square(2)$}
\put(12.5,4){$Square(3)$}

\put(0,2){\circle*{0.2}} \put(0,3){\circle*{0.2}}
\put(2,2){\circle*{0.2}} \put(2,3){\circle*{0.2}}

\put(6,2){\circle*{0.2}} \put(6,3){\circle*{0.2}}
\put(8,2){\circle*{0.2}} \put(8,3){\circle*{0.2}}

\put(12,2){\circle*{0.2}} \put(13,1){\circle*{0.2}}
\put(14,2){\circle*{0.2}} \put(13,3){\circle*{0.2}}

\put(0,2){\line(1,0){2}} \put(0,2){\line(2,1){2}}
\put(0,3){\line(1,0){2}} \put(0,3){\line(2,-1){2}}

\put(6,2){\line(1,0){2}} \put(6,2){\line(0,1){1}}
\put(6,3){\line(1,0){2}} \put(8,2){\line(0,1){1}}

\put(12,2){\line(1,-1){1}} \put(12,2){\line(1,1){1}}
\put(13,3){\line(1,-1){1}} \put(13,1){\line(1,1){1}}

\put(0,2.5){\oval(1.8,2)} \put(2,2.5){\oval(1.8,2)}
\put(6,2.5){\oval(1.8,2)} \put(8,2.5){\oval(1.8,2)}
\put(12,2){\oval(1,3)} \put(13.5,2){\oval(1.5,3)}

\put(6,0){$\alpha(G)=\frac{2}{2}=1$}
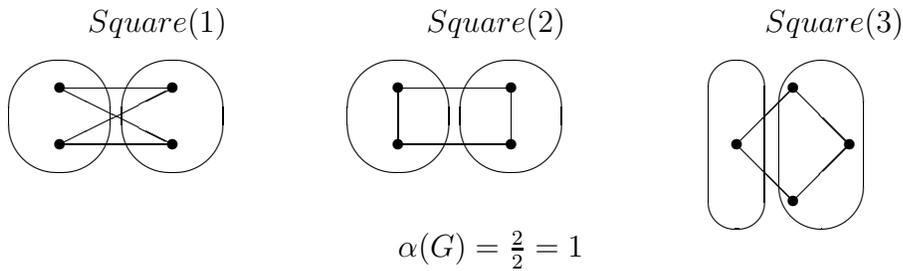
\put(0,-1){\footnotesize  FIG X: All kinds of partitioning in Square graph. The conductance is for partitioning $Square(2)$}
\put(0,-2){\footnotesize  The entanglement entropies are $S(\rho_{Square(1)})>S(\rho_{Square(2)})>S(\rho_{Square(3)})$}
\end{picture}
\newline
\newline
\newline
The amount of entanglement entropies in the examples of this section show that the entanglement entropy has direction relation with the number of connections between two parts. When two kinds of partitions have the same number of connections between two parts, the entanglement entropy has direction relation with the number of Schmidt numbers for those partitions.

\section{Investigation of entanglement entropy in the limit of large coupling and large size of the system}
\subsection{entanglement entropy in the limit of large size of the system}

We want to study the behavior of entropy of entanglement in (4.33), when $m_2$ and $n_2$ $\rightarrow \infty$.

So
\begin{equation}
lim_{m_2,n_2\rightarrow\infty} d=1
\end{equation}

Therefore the parameter $\nu \rightarrow \infty$ and finally the entropy $S(\rho)\rightarrow \infty$.

Also we are interested in the case that $m_1$ and $n_1$ tends to infinite.

\begin{equation}
lim_{m_1,n_1\rightarrow\infty} d=0
\end{equation}
In this case, the parameter $\nu \rightarrow 1$, so the entropy of entanglement tends to zero.

Suppose that the total size of the system tends to infinite, So $m_1,m_2,n_1,n_2\rightarrow \infty$

In this case $d\simeq \frac{2g\sqrt{m_2n_2}}{\sqrt{2gn_2}\sqrt{2gm_2}}\rightarrow 1$. Therefore the parameter $\nu \rightarrow \infty$ and also $S(\rho)\rightarrow \infty$.
\subsection{Large coupling limit}

We want to investigate large coupling limit in (4-33). So we can write
$$\frac{4g^{2}n_{1}n_{2}}{1+2gn_{2}}\simeq2gn_{1}(1-\frac{1}{2gn_{2}})$$
and
$$\frac{4g^{2}m_{1}m_{2}}{1+2gm_{2}}\simeq2gm_{1}(1-\frac{1}{2gm_{2}})$$
So
\begin{equation}
d=\frac{1}{\sqrt{1+\frac{1+\frac{n_{1}}{n_{2}}}{2gm_{2}}}\sqrt{1+\frac{1+\frac{m_{1}}{m_{2}}}{2gn_{2}}}}\cong1-\frac{1}{2}\varepsilon
\end{equation}
where
$$\varepsilon=\frac{1+\frac{n_{1}}{n_{2}}}{2gm_{2}}+\frac{1+\frac{m_{1}}{m_{2}}}{2gn_{2}}=\frac{N}{4gm_{2}n_{2}}$$
By using definition $(3-19)$, we have
\begin{equation}
\upsilon=\frac{1}{\sqrt{1-(1-\frac{1}{2}\varepsilon)^{2}}}\simeq\frac{1}{\sqrt{\varepsilon}}
\end{equation}
The entanglement entropy from (3-15) is
\begin{equation}
S(\rho)=\frac{\upsilon}{2}(1+\frac{1}{\upsilon})\log\frac{\upsilon}{2}(1+\frac{1}{\upsilon})-\frac{\upsilon}{2}(1-\frac{1}{\upsilon})\log\frac{\upsilon}{2}(1-\frac{1}{\upsilon})
\end{equation}
$$=\frac{1}{2}((\upsilon+1)(\log\frac{\upsilon}{2}+\frac{1}{\upsilon}))-\frac{1}{2}((\upsilon-1)(\log\frac{\upsilon}{2}-\frac{1}{\upsilon}))$$
So
\begin{equation}
S(\rho)=\log\frac{\upsilon}{2}+1=\frac{1}{2}\log\frac{gm_{2}n_{2}}{N}+1
\end{equation}
Where $m_{2}n_{2}$ is the size of the boundary and $N$ is the size
of the system.

\section{A special case that one vertex is in the first subset and the other vertices are in the second subset. Comparison of our result with the previous works}
Cardillo et al in [16] introduced the use of entanglement entropy
as a tool for studying the amount of information stored in quantum
complex networks.

By considering the ground state of a network of coupled quantum
harmonic oscillators, they computed the information that each node
has on the rest of the system. To this aim, they considered the
partition of the network into a node, say $i$, and its complement
$i_c$, i.e. the rest of the network. In their paper, they
asserted that the marginal entropies for $i$ and $i_c$ read:
\begin{equation}
S_i=S_{i_c}=(\mu_i+1/2)log(\mu_i+1/2)-(\mu_i-1/2)log(\mu_i-1/2)
\end{equation}
which they show that the parameter $\mu_i$ is
\begin{equation}
\mu_i^2=\frac{1}{4}\sum_{j,j'}S_{ij}^2 S_{ij'}^2
(\frac{1+2g\lambda_j}{1+2g\lambda_j'})^{1/2}
\end{equation}
where ${\lambda_j}$ are the eigenvalues of the network Laplacian
$L$ and matrix $S$ accounts for the normal mode transformation
that diagonalizes the network Laplacian: $L_d = S^TLS$ with $S^T S
= I$. It's clear that the matrix $S$ is not unique, So the above
equation for entropy has not a unique amount. It is easy to show
that the above equation for the entropy may be written in terms
of Laplacian:
\begin{equation}
\mu_i^2=\frac{1}{4}(1+2gL)_{ii}((1+2gL)^{-1})_{ii}=\frac{1}{4}V_{ii}V^{-1}_{ii}
\end{equation}
where
\begin{equation}
V=I+2gL=SDS^T=\left(\begin{array}{cc}
          A& B\\
            B^T& C\\
          \end{array}\right)
\end{equation}
is the potential of system, $A$ is a scalar(number), $B$ is a
column matrix and finally $C$ is a $n\times n$ matrix. We can
write: $V_{11}=(1+2gL)_{11}=1+2gd_1$

from the Schur complement it's clear that
$$V^{-1}_{11}=(A-BC^{-1}B^T)^{-1}=\frac{1}{(1+2gd_1)-BC^{-1}B^T}$$
\begin{equation}
\mu_1^2=\frac{1}{4}\frac{1+2gd_1}{(1+2gd_1)-BC^{-1}B^T}
\end{equation}

In the our method, we apply three stage to compute the
entanglement entropy. We started with  In order to diagonalize the
$I+2gL$, we should apply the local rotations $O_A$ and $O_C$ to
the system. But the matrix $A$ is a number, so the $O_A=I$. after
stage $1$ the matrix $V$ will be
\begin{equation}
\left(\begin{array}{cc}
          A& BO_C\\
            O_C^{\dagger}B^T& D_C\\
          \end{array}\right)
\end{equation}
Then the matrices $A$ and $C$ must transform to Identity. So after
transformation the $V$, will be
\begin{equation}
\left(\begin{array}{cc}
          I& D_A^{-1/2}B O_C D_C^{-1/2}\\
            (D_A^{-1/2} B O_C^{\dagger}D_C^{-1/2})^T& I\\
          \end{array}\right)
\end{equation}
By calculation of the singular value decomposition of column
matrix, in this stage we have
\begin{equation}
\left(\begin{array}{ccccc}
          I& \begin{array}{cccc}\frac{1}{\sqrt{A_{11}}}\|B O_C D_C^{-1/2}\| & 0 & \ldots & 0 \end{array} \\
            \begin{array}{c}\frac{1}{\sqrt{A_{11}}}\|(B O_C D_C^{-1/2})^T\| \\ 0 \\ \ldots \\ 0 \end{array}& I\\
          \end{array}\right)
\end{equation}
where $\|.\|$ is a kind of norm. So the parameter $\nu$ is
\begin{equation}
\nu_i^2=\frac{1}{1-\frac{B O_C
D_C^{-1/2}D_C^{-1/2}O_C^TB^T}{A_{11}}}=\frac{A_{11}}{A_{11}-BC^{-1}B^T}=\frac{1+2gd_1}{(1+2gd_1)-BC^{-1}B^T}
\end{equation}
So the relation between $\mu$ and $\nu$ is
$$\nu=2\mu$$
and the entropy is the same with (7-65)
\begin{equation}
S(\rho)=\frac{\nu +1}{2} log(\frac{\nu +1}{2})-\frac{\nu
-1}{2}log(\frac{\nu -1}{2})
\end{equation}

\section{Conclusion}
The entanglement entropy is obtained between two parts in
the quantum networks that their nodes are considered as quantum
harmonic oscillators.
The generalized Schur complement method is used to
calculate the Schmidt numbers and entanglement entropy between two
parts of graph.
\\Analytically, the Schmidt number can be calculated for special
graphs which their connections between parts are
complete, i.e. all graphs which their connections between each two parts are complete, have the same entropy (They may have every kinds of connections into each subsets). We proved it in theorem I. Some examples are given that the entanglement entropy and Schmidt number are calculated analytically in there.
More, the relationship between size of the
boundary of strata and entanglement entropy is obtained in the
limit of large coupling. Also the behavior of entanglement entropy is investigated when the size of system tends infinite.
\\ We investigated the conductance of graph and
entanglement entropy for some kinds of graphs.
\\Finally, we compare the results of our method with the results of
previous works in special case that one vertex is in the first
subset and the other vertices are in the second subset.
\\One expects that the entanglement entropy and Schmidt numbers can be used for distinguishing non-isomorphism
in two cospectral graphs.


\begin{thebibliography}{99}
\bibitem{1} A. K. Ekert, Phys. Rev. Lett. 67, 661 (1991).
\bibitem{2} C. H. Bennett, G. Brassard, C. Crépeau, R. Jozsa, A. Peres, W. K. Wootters,  Phys. Rev. Lett. 70, 1895 (1993).
\bibitem{3} R. Raussendorf, H. J. Briegel, Phys Rev. Lett. 86, 5188 (2001).
\bibitem{4} R. Horodecki, P. Horodecki, M. Horodecki, K. Horodecki, Rev. Mod. Phys. 81:865-942, (2009).
\bibitem{5} M. B. Plenio, S. Virmani, Quantum Inf. Comput. 7, 1 (2007).
\bibitem{6} M. J. Donald, M. Horodecki, and O. Rudolph, J. Math. Phys. 43, 4252–4272 (2002).
\bibitem{7} J. Sperling, W. Vogel, Phys. Scr. 83, 045002 (2011).
\bibitem{8} Y. Guo and H. Fan, Quant-ph: 1304.1950 (2013).
\bibitem{9} L. Amico, R. Fazio, A. Osterloh, V. Vedral, Rev. Mod. Phys. 80, 517-576 (2008).
\bibitem{10} J. M. Matera, R. Rossignoli, and N. Canosa, Phys. Rev. A 86, 062324 (2012).
\bibitem{11} G. Adesso, S. Ragy, A. R. Lee, Open Syst. Inf. Dyn. 21, 1440001 (2014).
\bibitem{12} O. Cernotík and J. Fiurášek, Phys. Rev. A 89, 042331 (2014).
\bibitem{13} G. Adesso and S. Piano, Phys. Rev. Lett 112, 010401 (2014).
\bibitem{14} F. Nicacio and M. C. de Oliveira, Phys. Rev. A 89, 012336 (2014).
\bibitem{15} D. Buono, G. Nocerino, S. Solimeno and A. Porzio, Laser Phys. 24 074008 (2014).
\bibitem{16} A. Cardillo, F. Galve, D. Zueco, J. G. Gardenes, Phys. Rev. A 87, 052312 (2013).
\bibitem{17} F. Zhang, The Schur Complement and Its Applications, Springer, (2005).




\end{thebibliography}
\end{document}